\documentclass[twocolumn]{aastex631}

\usepackage{braket}
\usepackage{graphicx}
\usepackage[normalem]{ulem}
\usepackage{float}
\begin{document}

\title{First detection of Circular Polarization in radio continuum towards a Massive Protostar}

\author[0000-0001-9716-5319]{A.G. Cheriyan}
\email{Corresponding author: A. G. Cheriyan, S. Vig\\amalgeorgecheriyan@gmail.com\\sarita@iist.ac.in}
\affiliation{Department of Earth and Space Sciences \\
Indian Institute of Space Science and Technology \\
 Thiruvananthapuram 695547, India}

\author[0000-0002-3477-6021]{S. Vig}
\affiliation{Department of Earth and Space Sciences \\
Indian Institute of Space Science and Technology \\
 Thiruvananthapuram 695547, India}


\author[0000-0001-9829-7727]{Nirupam Roy}
\affiliation{Department of Physics \\
Indian Institute of Science \\
Bangalore 560012, India}

\author[0000-0001-9371-7104]{Samir Mandal}
\affiliation{Department of Earth and Space Sciences \\
Indian Institute of Space Science and Technology \\
 Thiruvananthapuram 695547, India}

 \author[0000-0003-2862-5363]{C. Carrasco-Gonz\'{a}lez}
\affiliation{Instituto de Radiastronom\'{i}a y Astrof\'{i}sica (IRyA)\\
Universidad Nacional Aut\'{o}noma de M\'{e}xico (UNAM)\\
Morelia, Michoac\'{a}n, M\'{e}xico}

\author[0000-0002-4731-4934]{A. Rodr\'{i}guez-Kamenetzky}
\affiliation{Instituto de Astronom\'{i}a Te\'{o}rica y Experimental (IATE)\\
Universidad Nacional de C\'{o}rdoba (UNC)\\
C\'{o}rdoba, Argentina}

 \author[0000-0003-1933-4636]{A. Pasetto}
\affiliation{Instituto de Radiastronom\'{i}a y Astrof\'{i}sica (IRyA)\\
Universidad Nacional Aut\'{o}noma de M\'{e}xico (UNAM)\\
Morelia, Michoac\'{a}n, M\'{e}xico}

\begin{abstract}

Polarization measurements provide strong constraints on magnetic fields in star-forming systems. While magnetic field estimates of a few kiloGauss (kG) have been obtained near the surface of low-mass protostars, there are no analogous measurements in the immediate vicinity of the surface of massive protostars. We report the measurement of radio continuum circular polarization (CP) towards a massive protostar IRAS~18162-2048 for the first time wielding Karl G. Jansky Very Large Array (VLA) observations. The fractional CP varies between $3-5\%$ across the observed frequency range of $4-6$~GHz. We consider multiple hypotheses for the production of CP and propose (i) gyrosynchrotron emission and (ii) Faraday conversion due to turbulence in the magnetic medium - both driven by mildly relativistic electrons as plausible mechanisms.  We estimate, for the first time, a magnetic field $B\gtrsim20-35$~G close to the massive protostar. The Lorentz factor of the low energy electrons is estimated to be in the range $\gamma_{min}\sim5-7$ for gyrosynchrotron emission and $80-100$ for Faraday conversion from our observations. The magnetic field estimate can provide important constraints to the formation models of massive stars. 

\end{abstract}

\keywords{Star formation (1569) --- Protostars (1302) --- Radio astronomy (1338) --- Polarimetry (1278) --- Interstellar magnetic fields (845) --- Interstellar synchrotron emission (856)}

\section{Introduction} \label{sec:Intro}

The influence of magnetic fields in the formation of massive stars remains a topic of ongoing scientific debate \citep{2010HiA....15..438C,2014ApJ...792..116Z}. Observations and models suggest that magnetic fields play a crucial role in launching powerful jets and outflows, enhancing the degree of outflow collimation, stabilizing both Keplerian and sub-Keplerian disks against fragmentation \citep{2011MNRAS.417.1054S,2012MNRAS.422..347S}, as well as transporting excess angular momentum out of the system \citep{2011ARA&A..49..195A}. Towards low-mass protostars, surface magnetic fields up to several kiloGauss have been established enabling robust models of magnetic field assisted accretion of material onto the protostar \citep{1999A&A...341..768G,1999ARA&A..37..363F}. On the other hand, little is known about magnetic fields in the close environments of massive protostars, and magnetic field estimates have only been obtained in the dense cores and/or disk-jet region using dust polarization measurements or polarization measurements employing masers \citep{2006A&A...448..597V,2014ApJ...794L..18Q,2023A&A...680A.107M}. The linear polarization of thermal lines, such as those from CO and SiO \citep{1981ApJ...243L..75G} can also serve as a probe of magnetic fields in the plane of the sky. Nonetheless, detections to date have been limited and remain marginal \citep{2021ApJ...922..139T}.

Circular polarization (CP) measurements - both in radio continuum as well as optical Doppler-shifted lines have been employed as one of the tools to estimate the magnetic field environment close to the surface of low-mass protostars \citep{1997A&A...326.1135D,2015ApJ...801...91D}. 
The CP in radio continuum has been attributed to gyrosynchrotron emission from mildly relativistic electrons associated with the coronal magnetic field of low-mass YSOs \citep{2015ApJ...801...91D}. However, no analogous measurements are available for massive protostars. Radio CP has also been detected towards AGN, with jets speculated as the plausible origin \citep{2001ApJ...560L.123B}. The fraction of circularly polarised light towards low-mass YSO cores have been found to be as high as $\sim70\%$ \citep{2015ApJ...801...91D} whereas the CP towards AGN cores has been found to be of the order of a few percent \citep{2004ApJ...602L..13H}. Observations indicate a helical magnetic field morphology for astrophysical jets in both the classes of objects, suggesting a universal mechanism for their launch and propagation \citep{2018Galax...7....5G, 2021ApJ...923L...5P,2022ApJ...927L..27L}. The intriguing kinship of (i) low and high mass star-formation, and (ii) similar jet-launching mechanisms in a variety of astrophysical objects leads us to ponder whether CP could also be observed towards massive protostars.

In this work, we examine the radio continuum polarization towards IRAS~18162-2048, (hereafter I18162) which is associated with a massive protostar, equivalent to ZAMS type B0. I18162 is located at a distance of about 1.4 kpc \citep{2020ApJ...888...41A}, and drives the largest known, highly collimated and most luminous jet in our Galaxy - the HH80-81 jet \citep{1989RMxAA..17...59R,1993ApJ...416..208M,2023JApA...44...57M,2025A&A...695A..11M}. Millimeter wavelength observations towards I18162 have revealed twenty-five cores - with the most massive MM1 core responsible for the HH80-81 jet \citep{2019A&A...623L...8B}. Synchrotron emission has been detected in the inner lobes ($\sim0.4$~pc) of the jet through linear polarization (LP) measurements \citep{2010Sci...330.1209C}, and suggests a helical configuration of the magnetic field \citep{2025ApJ...978L..31R}. 
 I18162 is an ideal target for the investigation of CP in massive protostars due to the presence of a large-scale massive magnetized jet, availability of radio observations across multiple epochs and frequencies, as well as its relative proximity. 

In order to examine the polarization properties towards I18162 and the associated jet, we have conducted full Stokes radio observations using the Karl G. Jansky Very Large Array (VLA) between 4-6~GHz. The observations were carried out as two separate runs on 20 and 21~Dec~2018. The data from both runs were concatenated to generate images in all four Stokes parameters (I, Q, U, V), enabling a comprehensive analysis of linear and circular polarization. These polarization observations provide insights into the magnetic field close to the massive protostar. We investigate plausible emission mechanisms, as well as propagation effects in the immediate ambient medium in order to understand the results. 

The structure of the paper unfolds as follows: Section \ref{sec:Observations} provides information about the observational techniques and subsequent data reduction processes. Our findings are presented in Section \ref{sec:Results}, while Section \ref{sec:Discussion} delves into the discussion of results. Section \ref{sec:Summary} provides a succinct overview of the conclusions from our study.

\begin{figure*}
    \centering
    \includegraphics[width=0.9\linewidth]{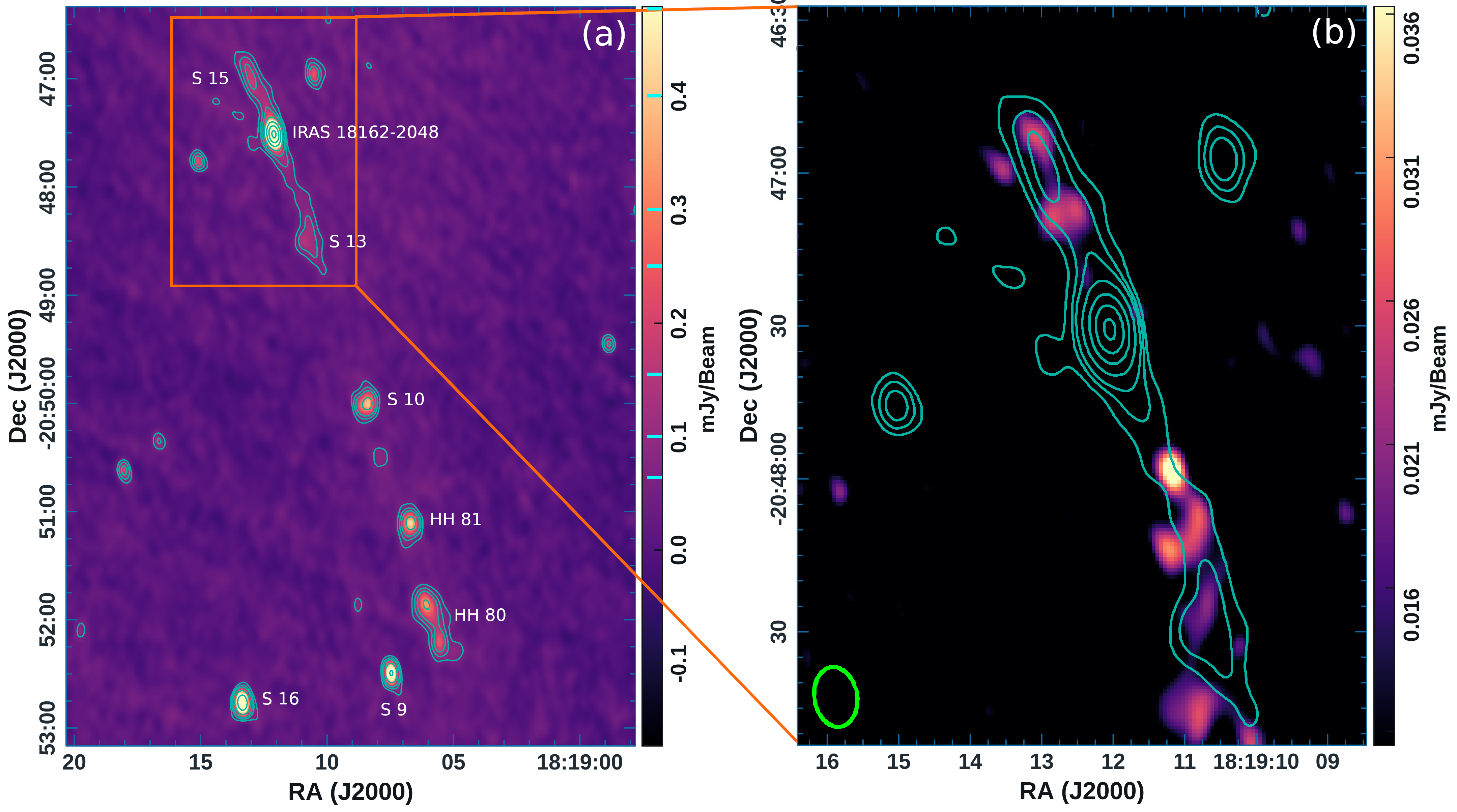}
    \caption{(a): Total intensity radio map of I18162 region at VLA C band. The contour levels in cyan are marked in the color bar, where beam: $7.5'' \times 5.0''$. (b): Linearly polarized emission observed towards the near jet lobes of I18162. The cyan contours overlaid on the image are the Stokes $I$ contours which are the same as on (a). The green ellipse at the bottom left corner shows the beam size. The color bar is shown on the right side of the image.}
    \label{fig:comb}
\end{figure*}


\section{Observation and Data Reduction} \label{sec:Observations}

Low radio frequency observations of the protostar, I18162, have been carried out in the C band using the Karl G. Jansky Very Large Array (VLA) of the National Radio Astronomy Observatory with circular feeds (Project ID: 18B-029). Observations were conducted in two separate runs on 21st and 22nd December 2018,  with a total on-source time of approximately 8 hours (roughly 4 hours in each run). The phase centre for these observations was located at $\alpha(J2000) = 18^{h}~19^{m}~06.4^{s}$ and $\delta(J2000) = -20^{\circ}~51'~32.0''$. For both observational runs, the phase calibrator J1911-2006 was employed. J2355+4950 was utilized as the polarization calibrator. The flux and bandpass calibrations for both runs were performed using 3C286. Each measurement set comprised 32 spectral windows, each consisting of 64 channels, leading to a total bandwidth of 128~MHz.

Data reduction and imaging were carried out utilizing the Common Astronomy Software Applications (CASA) package \citep{2022PASP..134k4501C}. 
For polarization calibration, observations of 3C286 were utilized to determine the absolute polarization angle, while observations of the polarization calibrator J2355+4950 were leveraged to identify and rectify antenna-based leakage terms accountable for instrumental polarization across each spectral window. Notably, the polarization calibrator exhibits stable flux and minimal polarization at the given frequencies, facilitating the correction of leakage terms within a single scan. To assess the quality of our polarization calibration, we measured the polarization angle and degree of the flux and phase calibrators employed. Our results matched with the values listed in the VLA Polarization Database \citep{1982AJ.....87..859P}. 

Rigorous checks were carried out to detect and address corrupted data in the spectral windows - stemming from factors such as radio frequency interference, non-operational antennas, and their time dependence. The source data concatenated from the two-day observations were imaged in ten clean spectral windows individually, using the TCLEAN multi-scale deconvolution task with $nterm = 2$ to obtain all four Stokes images. We then used these images to perform phase-only self-calibration and imaged in Stokes \textit{I}, \textit{Q}, \textit{U} and \textit{V}, applying the solutions to the respective Stokes data. We applied a uv taper of $20~k\lambda$ to the images for consistency checks with LP measurements of the jet in the earlier images \citep{2010Sci...330.1209C}, and primary beam correction has been carried out to all images. This leads to a synthesized beam size of $9.5'' \times 6''$ with a position angle of $+2^{\circ}$. The final Stokes images have an average rms of $10-20$~$\mu$Jy beam$^{-1}$. The linearly polarized image is made from the Stokes $Q$ and $U$ images as $LP =~\sqrt{Q^{2} + U^{2}}$, whereas the Stokes $V$ image is used to generate the circularly polarized image, which is related to the difference between the left and right-handed circular polarization, correlated visibilities. In order to ensure the robustness of our results, we have carried out multiple checks, the details of which are given in the Appendix \ref{appendix}.


\begin{figure*}
    \centering
    \includegraphics[width=0.9\linewidth]{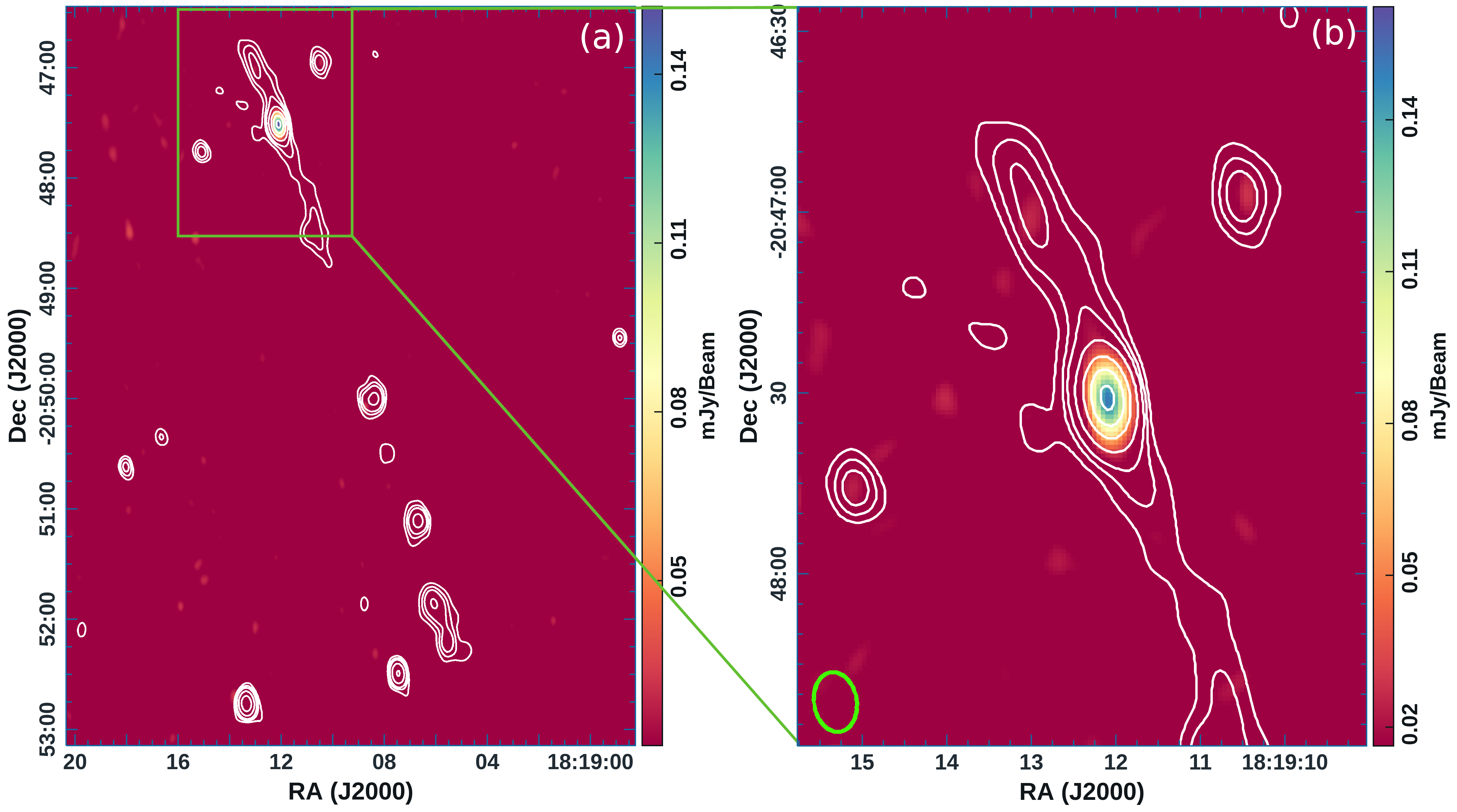}
    \caption{(a): Circularly polarized Stokes $V$ emission observed towards the central source I18162. The white contours overlaid on the image are the Stokes $I$ contours, which are the same as in Fig.~\ref{fig:comb}. The color bar is shown on the right side of the image. (b): Zoomed-in view of the central region exhibiting the circularly polarized emission. The green ellipse at the bottom left corner shows the beam size.}
    \label{fig:stokesV}
\end{figure*}
%

\section{Results} \label{sec:Results}

\subsection{Polarised emission}

The total intensity ($I$) radio continuum image of I18162 shows the protostar and the bipolar radio jet with various emission knots, including Herbig-Haro objects HH80 and HH81. This emission, shown in Fig.~\ref{fig:comb}(a), traces the ionized emission from the protostar and the jet. The brightest emission is observed towards the massive protostar I18162. The source names displayed in Fig.~\ref{fig:comb}(a) are taken from \citet{1993ApJ...416..208M}. The linearly polarized emission detected from this region can be viewed in Fig.~\ref{fig:comb}(b). We find linearly polarised emission towards the inner jet that includes the lobes S~15 and S~13. Notably, we do not detect LP towards I18162. This is consistent with the emission detected earlier \citep{2010Sci...330.1209C} and in recent times \citep{2025ApJ...978L..31R} with similar rms in the images.

The CP is measured through the Stokes parameter $V$, and for our region of interest - this is shown in Fig~\ref{fig:stokesV}. We find CP at the level of $\sim80$~$\mu$Jy at the level of about $10\sigma$ ($\sigma$ is the rms noise in the image) towards I18162 but do not find any emission towards the jet lobes. The emission towards I18162 is the first detection of CP in radio continuum towards a massive protostar. The emission is compact in nature, suggesting emission from the region in the close vicinity of the protostar, $\lesssim 1000$~au (corresponding to the beam size). 

\subsection{Spectral index and Polarization fraction}

\begin{figure}[ht]
\includegraphics[width=1\columnwidth]{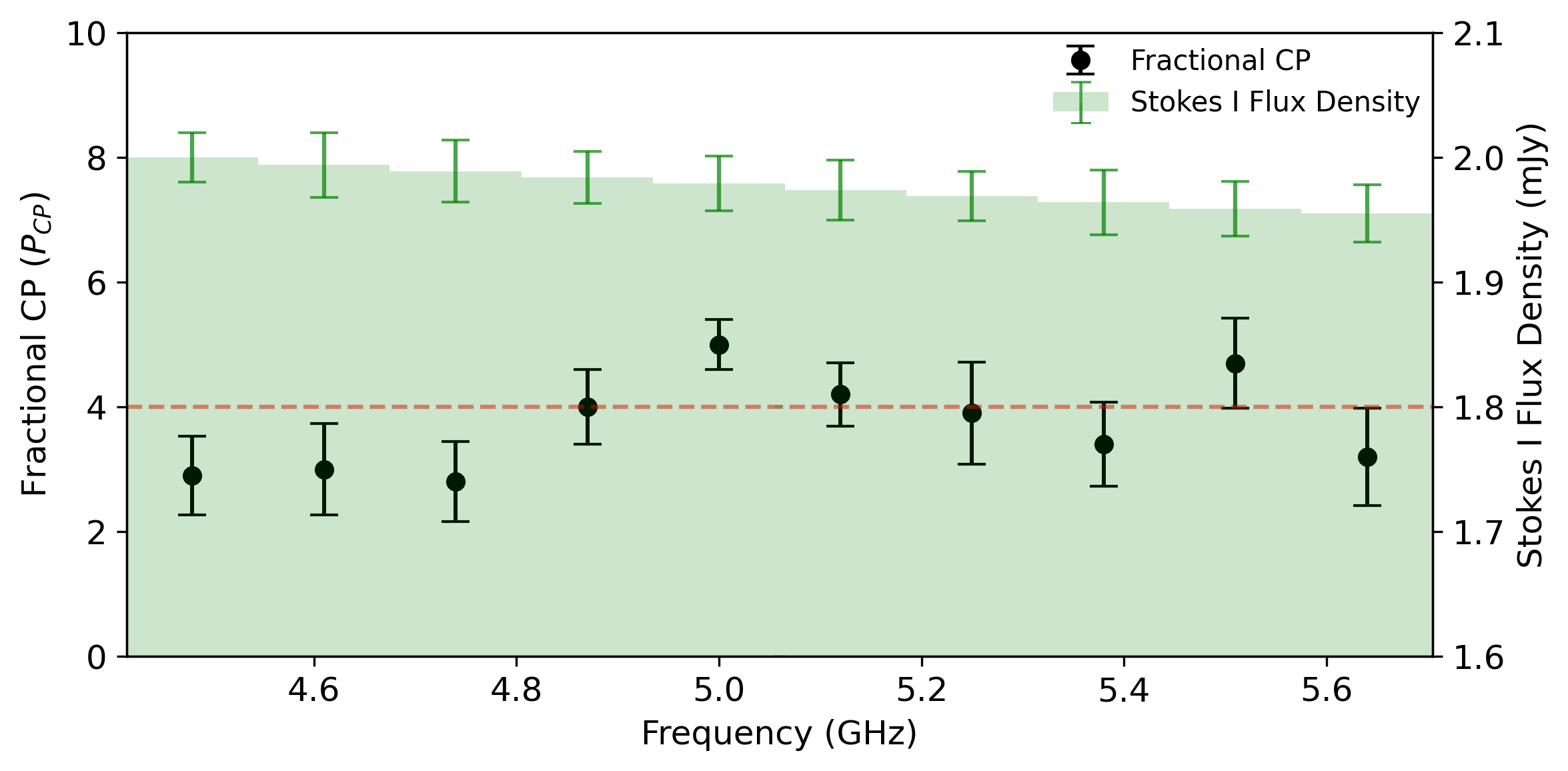}
\caption{Plot of fractional circular polarization (solid black circles) and Stokes~$I$ flux densities (histogram in green) as a function of frequency. The data depict the fractional circular polarization at the central frequencies of spectral windows used in the analysis. The red dashed line represents the nominal fractional CP of $4\%$ assumed in our estimations; see Sec.~3 for details.}
\label{fig:pol_frac}
\end{figure}

 The total flux densities as a function of frequency can be utilised in estimating the in-band spectral index of I18162 between 4-6~GHz. This, in turn, can be used to surmise about the emission mechanism(s) at work, thus providing significant insights into the properties of the region from where the radio emission is emanating. The flux densities of I18162 are estimated by integrating emission from within a common elliptical aperture in all the images. The total (i.e. Stokes $I$) flux densities are shown as a histogram in Fig~\ref{fig:pol_frac}.  By comparing these flux densities, we find that I18162 has a relatively flat spectral index $\alpha \sim -0.10\pm0.02$. Earlier measurements of spectral index obtained towards I18162 range from flat to mildly positive, and lie in the range $0.02-0.3$ \citep{2017ApJ...851...16R,2018MNRAS.474.3808V,2023JApA...44...57M}. This range of spectral indices covers different frequency ranges and epochs. The spectral index values have been explained as a combination of synchrotron emission from the jet and the thermal free-free emission emanating from the surrounding H II region \citep{2018MNRAS.474.3808V,2012ApJ...758L..10M,2023JApA...44...57M}, in addition to partial optical thickness of the jet \citep{2017ApJ...851...16R}. 
 It is interesting to note that AGN cores with jets as well as low mass YSOs displaying CP also exhibit relatively flat spectral indices ($\pm 0.2$), which is similar to our results \citep{2000MNRAS.319..484R,2006A&A...453..959M}. This is discussed further in the next section. The circular polarization fraction, calculated as $P_{CP} = V/I$, varies between $3-5~\%$ and does not exhibit a discernible trend with frequency within the observed bandwidth; this is illustrated in Fig.~\ref{fig:pol_frac}.


\section{Discussion} \label{sec:Discussion}

Several mechanisms have been proposed to explain the origin of CP towards low-mass YSOs \citep{1985ARA&A..23..169D,1996AJ....111..918P,1999ARA&A..37..363F}, as well as AGNs and X-ray binaries \citep{2001ASPC..250..152W}. However, only a limited number of observations are available towards these sources due to variability of CP, low fractional polarization, as well as relative difficulties in carrying out polarization observations. The mechanisms suggested for the generation of CP include the (i) coherent radiation mechanism, (ii) scintillation-induced CP, (iii) gyrosynchrotron mechanism, and (iv) Faraday conversion. We investigate these mechanisms with respect to the case of I18162.

\subsection{Coherent Radiation mechanism}\label{sec:4.1}

Coherence from maser action or particle bunching typically produces a narrow-band signal, with high circular polarization (CP) confined typically to about 100-400 MHz \citep{2000MNRAS.317..497B}. This mechanism has been proposed to explain the high CP observed in intra-day variable sources such as AGN cores \citep{1995A&A...301..641R}. However, the CP spectrum observed towards I18162 is flat across the bandwidth observed, suggesting that this is not a likely origin. This emission mechanism can, therefore, be excluded from our consideration.

\subsection{Scintillation induced CP}\label{sec:4.2}

An alternative mechanism proposed is through the scintillation of the magnetized, inhomogeneous plasma medium that induces CP, even if the source is inherently unpolarized \citep{2000ApJ...545..798M}. This mechanism predicts a steep spectrum and a high ratio of linear-to-circular polarization (LP/CP), i.e. $R_{LC} >> 1$ \citep{2000ApJ...545..798M,2001ApJ...560L.123B}. However, the flat CP spectrum and absence of LP in images indicate $R_{LC} \lesssim 0.3$ when we employ the upper limit of LP detection. This suggests that interstellar propagation is not responsible for CP.

\subsection{Gyrosynchrotron mechanism}\label{sec:4.3}

We consider the gyrosynchrotron radiation phenomenon \citep{1968ApJ...154..499L, 1985ARA&A..23..169D} arising from mildly relativistic electrons as a potential mechanism of CP production. The emission from individual electrons is circularly polarized when the line-of-sight is close to the direction of the magnetic field for electrons streaming perpendicular to the magnetic field lines. This cancels out for pure electron-positron isotropic distributions and is enhanced for anisotropic electron distributions preferentially perpendicular to the local magnetic field \citep{2002A&A...388.1106B}. In a uniform magnetic field $B$, the maximum fractional CP, $P_{CP,max}$ at a frequency $\nu$ is given by \citep{1971Ap&SS..12..172M}

\begin{equation}
\label{eq:1}
P_{CP,max} = \frac{\cot\theta}{3}\left[\frac{\nu}{3\nu_{B}\sin\theta}\right]^{-1/2}f(\alpha)
\end{equation} 
\noindent Here, $\theta$ is the inclination angle of the jet with respect to the observer, and $f(\alpha)$ is a weakly dependant function of spectral index $\alpha$, with values between 0.6 and 2 for $\alpha = 0$ to 2 \citep{1971Ap&SS..12..172M}, respectively. $\nu_{B} = eB/mc$ is the electron gyrofrequency in presence of magnetic field $B$, $e$ and $m$ are the charge and mass of electron, respectively, and $c$ is the speed of light. In the case of I18162, since $\alpha\sim -0.1$, i.e. flat spectral index, we take $f(\alpha) \sim 0.6$. Taking $P_{CP,max}\sim4\%$ at 5~GHz from our observational data, and considering an inclination in the range $\theta\sim 34^\circ-45^\circ$ \citep{1998AJ....116.1940H,2020ApJ...888...41A}, the magnetic field strength is estimated to be in the range $B \sim 20-35$~G. The Lorentz factor of the low-energy electrons giving rise to gyrosynchrotron emission can be calculated by the expression $\gamma = 0.1915\,(\cot\theta)\times P_{CP}^{-1}$ \citep{1999ApJ...526L..85S}. This gives $\gamma\sim 5-7$, consistent with the expected  Lorentz factor of $\gamma < 10$ for gyrosynchrotron emission \citep{2006A&A...453..959M}. This suggests that the gyrosynchrotron emission is a viable mechanism for producing the measured CP in I18162. 

\subsection{Faraday conversion}\label{sec:4.4}

The possibility of LP radiation getting converted to CP through Faraday conversion along the line of sight towards the source is also a likely hypothesis of the observed emission. The conversion is possible due to the presence of varying magnetic field directions as a result of turbulence along the line-of-sight \citep{{1977ApJ...214..522J},{1977ApJ...211..669H}}. The Faraday conversion and depolarization have been employed for modelling the observed CP and lack of LP from Sgr~A* \citep{1999AAS...195.6203B}. In this case, the spectral characteristics of CP can result in a variety of spectral indices, leading to flat or mildly positive spectral indices e.g. $\alpha\lesssim0.3$ observed towards AGN cores \citep{1977ApJ...214..522J}. The flat spectral index of CP towards I18162 renders this mechanism feasible for explaining the observed emission. The fractional CP induced by the turbulence in the magnetic medium is given by the expression \citep{2002A&A...388.1106B} below. 
\begin{equation}\label{eq:3}
P_{CP} \approx 0.5~\frac{s+1}{s+7/3} \left(\frac{1}{K_{out}R}\right) \frac{\tau_{C}}{\tau_{F}}~\frac{B_{z}}{B_{0}}~\cos\theta
\end{equation}
\noindent Here, $s$ is related to the electron spectral index as $s =2\alpha + 1$, $K_{out}$ is the inverse length of a single turbulent cell, while $K_{out}R$ represents the number of turbulent cells along the path of the ray traversing a distance $R$ in the medium. $B_{z}\cos\theta$ is the source magnetic field along the line of sight, $\theta$ is the inclination angle of the jet with respect to the line of sight, and $B_{0}$ is the sum of ordered field from source and turbulent field from ambient medium in the vicinity. Here $\tau_{C}$ and $\tau_{F}$ represent the Faraday conversion depth, and Faraday rotation depth expressed in terms of total optical depth  ($\tau$) by the following expressions.
\begin{equation}\label{eq:4}
\frac{\tau_{C}}{\tau} = \frac{2}{(s-2)K_{out}R} \left[\left(\frac{\gamma_{rad}}{\gamma_{min}}\right)^{s-2}-1\right] \end{equation}
\begin{equation}\label{eq:5}
\frac{\tau_{F}}{\tau} = \frac{2(s+2)}{(s+1)K_{out}R} \left(\frac{\gamma_{rad}}{\gamma_{min}}\right)^{s}~\frac{\ln (\gamma_{min})}{\gamma_{min}}   
\end{equation}
In this context, $\gamma_{min}$ denotes the minimum Lorentz factor for a power-law electron distribution while $\gamma_{rad}$ represents the Lorentz factor at which most of the emission occurs.


\begin{figure}[ht]
\includegraphics[width=1\columnwidth]{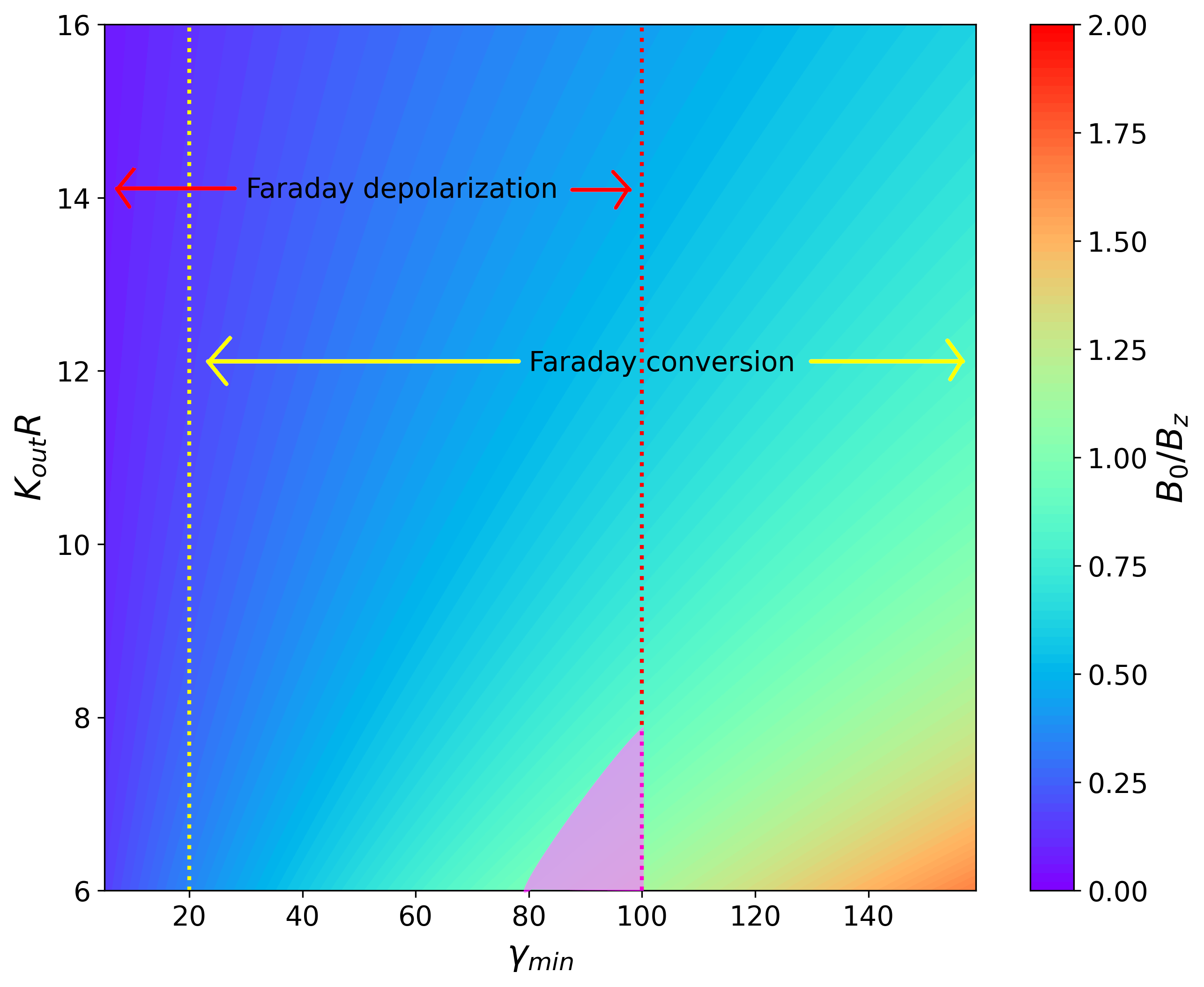}
\caption{Variation of the number of turbulent cells ($K_{out}R$) with respect to the change in the Lorentz factor ($\gamma_{min}$). The colour bar represents the ratio of the total magnetic field and magnetic field towards the line of sight ($B_{0}/B_{z}$). The pale magenta-coloured region represents the feasible values of $\gamma_{min}$ and $K_{out}R$ considering $B_{0} > B_{z}$, Faraday conversion and Faraday depolarization.}
\label{fig:koutR}
\end{figure}


For $P_{CP}\sim4\%$ with a flat spectrum $s \sim 1$, we consider the case when Faraday conversion becomes dominant, implying CP is higher than LP \citep{2002A&A...388.1106B}. In this situation, $\gamma_{min} > 20$ and $\gamma_{rad} \sim \gamma_{min}/0.9$. This is applicable to our case as LP is below the detection threshold. In contrast, for Faraday depolarization to be effective, the minimum Lorentz factor should be less than 100 \citep{2002A&A...388.1106B}, i.e. $\gamma_{min} < 100$. By taking $\theta \sim45^\circ$ \citep{2020ApJ...888...41A}, the ratio of the total magnetic field $B_{0}$ to the magnetic field associated with the protostar $B_{z}$ can be determined. We take the number of turbulent cells $K_{out}R \sim 6-16$, where the lower limit is decided by $\tau_{C}/\tau << 1$ when a significant fraction of LP is converted to CP. The possible values of $B_{0}/B_{z}$ is shown in colour as a function of $\gamma_{min}$ and $K_{out}R$ in Fig.~\ref{fig:koutR}. We find that within the constraints, $K_{out}R \sim 6-8$ and $\gamma_{min} \sim 80-100$, the observed CP fraction can be accounted for by this mechanism when $1\leq B_{0}/B_{z}\leq 1.25$. Thus, for a magnetic field of, say, $B_{z}\sim50$~G near the protostar, the total magnetic field, including the turbulent magnetic field along the line-of-sight, would be $B_{0}\sim50 - 62$~G.

The characteristic variability timescale for the stochastic variation of fractional CP in this hypothesis is given by,
\begin{equation}
   \Delta~t = \frac{R}{v_{A}(K_{out}R)} 
\end{equation}
Here, $v_{A}= B/\sqrt{4\pi\rho}$ is the Alfven velocity, and $\rho= n \mu m_{H}$, where $\mu$ is the mean mass of particle in units of mass of a proton $m_{H}$, with $n$ being the number density in the immediate proximity of the source \citep{2002A&A...388.1106B}. Taking a typical number density of $10^{7}$ cm$^{-3}$ \citep{2012A&A...542A..76H}, close to the massive protostar, $\mu$ to be $0.5$, and assuming a magnetic field $B\sim 50$~G the variability timescale is found to be of the order of 20 days. 

 
Thus, we note that either of the above mechanisms, Gyrosynchrotron emission, Faraday conversion or a combination of these, are plausible for the case of I18162. This study provides the first estimate of the magnetic field of $20-35$~G in the close vicinity of a massive protostar from CP measurements in radio continuum, assuming the origin to be gyrosynchrotron emission. We note that this mechanism is proposed and utilised to estimate the magnetic field around low mass YSOs \citep{2024arXiv240804025K} as well as AGN \citep{2021Galax...9...58G}. 
Towards intermediate mass, HAeBe pre-main sequence stars, magnetic fields of $10-250$~G have been measured \citep{2004ApJ...617..406M,2007MNRAS.376.1145W}. Towards low-mass YSOs, magnetic fields of several kiloGauss have been measured \citep{2007prpl.conf..479B,2022MNRAS.511..746W}.  For massive protostars, magnetic field measurements to date are mostly from high-density cores using dust polarization (up to several tens of mG) \citep{2007MNRAS.382..699C,2014ApJ...794L..18Q}, and masers from disk/jet region ($\sim0.1 - 0.6$~G)\citep{2006A&A...448..597V,2023A&A...680A.107M}. The magnetic field of $20-35$~G towards a massive protostar, presented here, likely represents the projection of the poloidal field close to the protostar. This magnetic field estimate would be a lower limit due to the possibility of CP being affected by depolarization from the surrounding medium along the line of sight. As there are no measurements of magnetic fields in the immediate proximity of massive protostars to draw comparisons, we explore an alternative approach. We examine magnetic fields observed towards certain massive stars. 

A small fraction of early-type OB stars display surface magnetic fields of several hundred to thousands of gauss \citep{2021MNRAS.501.2677D,2023Galax..11...40K}. With their convective cores and radiative envelopes, analyses show that dynamo processes prevalent in low-mass stars are unlikely to produce the observed fields in these magnetic stars \citep{2001ApJ...559.1094C}. Rather, the presence of remnant fossil fields from the early star formation processes are invoked to explain them \citep{2004Natur.431..819B,2015SSRv..191...77F}. 
The fossil field hypothesis suggests that the fields associated with magnetic stars were generated during an earlier phase of the star's evolution that has been retained through magnetic self-induction \citep{1999stma.book.....M}. According to this theory, magnetic fields from the interstellar medium permeate molecular clouds, and as these clouds undergo gravitational collapse into protostars, the fields are both advected and amplified. This process can generate magnetic fields up to a few hundred gauss in massive stars, which is consistent with the observed results \citep{2019EAS....82..345A}. Given the strength of the magnetic field estimated towards I18162, we speculate that I18162 could be a precursor to a massive magnetic star. The magnetic field estimate towards this massive protostar could provide important constraints to star-formation models. This will aid in a better understanding of the role of the magnetic field in massive star formation, such as the significance of magnetic field during the collapse process, the effect of magnetic fields on fragmentation and multiplicity \citep{2011ApJ...742L...9C} and formation of isolated massive stars \citep{2021ApJ...912..159P}. 

We end with a note that although we have speculated about several emission mechanisms to account for the observed CP in this work, more observations are required to determine the nature of the mechanism at work towards massive protostars.

%
\section{Summary} \label{sec:Summary}

We have carried out polarization study of the protostar I18162 using the VLA at $4-6$~GHz. For the first time, we have detected CP towards this protostar. The fractional CP lies in the range $3-5\%$, showing no perceptible variations in frequency. To explain the observed emission, we have considered several emission mechanisms that can account for the measured CP, and we have narrowed it down to two mechanisms. The gyrosynchrotron emission and the Faraday conversion of LP to CP, both due to mildly relativistic electron with $\gamma_{min}\sim5-7$, can explain the observed CP. Assuming that the gyrosynchrotron emission is responsible for the observed emission, we estimate the magnetic field near the protostar to be $B\sim20-35$~G. This is the first magnetic field measurement close to massive protostars utilizing CP measurements. In the case of Faraday conversion, the Lorentz factor is found to be between $\gamma_{min}\sim80-100$. These low-energy relativistic electrons can also reduce LP through Faraday depolarization and convert LP into CP by Faraday conversion, with turbulence in the magnetic field along the line of sight in the ambient medium. Based on our measurements, we have constrained the number of turbulent cells and find that this is likely to be between $K_{out}R\sim6-8$. The magnetic field obtained suggests that I18162 is a likely precursor to a magnetic massive star. However, more observations are needed to confirm this.

\section{Acknowledgments} \label{sec:Acknowledgments}

We thank the referee for carefully perusing the manuscript and giving inputs that have helped improve the paper. We thank the staff of Karl G. Jansky Very Large Array (VLA), who made the radio observations possible. The National Radio Astronomy Observatory is a facility of the National Science Foundation operated under a cooperative agreement by Associated Universities, Inc. A.G.-C. acknowledges the Indian Institute of Science (IISc), Bengaluru, where partial data analysis was carried out. C.C.-G, A.P and A.R.-K acknowledge support from UNAM DGAPA-PAPIIT grant IG101224 and from CONAHCyT Ciencia de Frontera project ID 86372.

\bibliography{sample631}{}
\bibliographystyle{aasjournal}
\clearpage
\appendix
\twocolumngrid
\restartappendixnumbering

\section{Checks carried out for confirmation of CP detection} \label{appendix}

\begin{figure}[H]
    \centering
    \includegraphics[width=\textwidth]{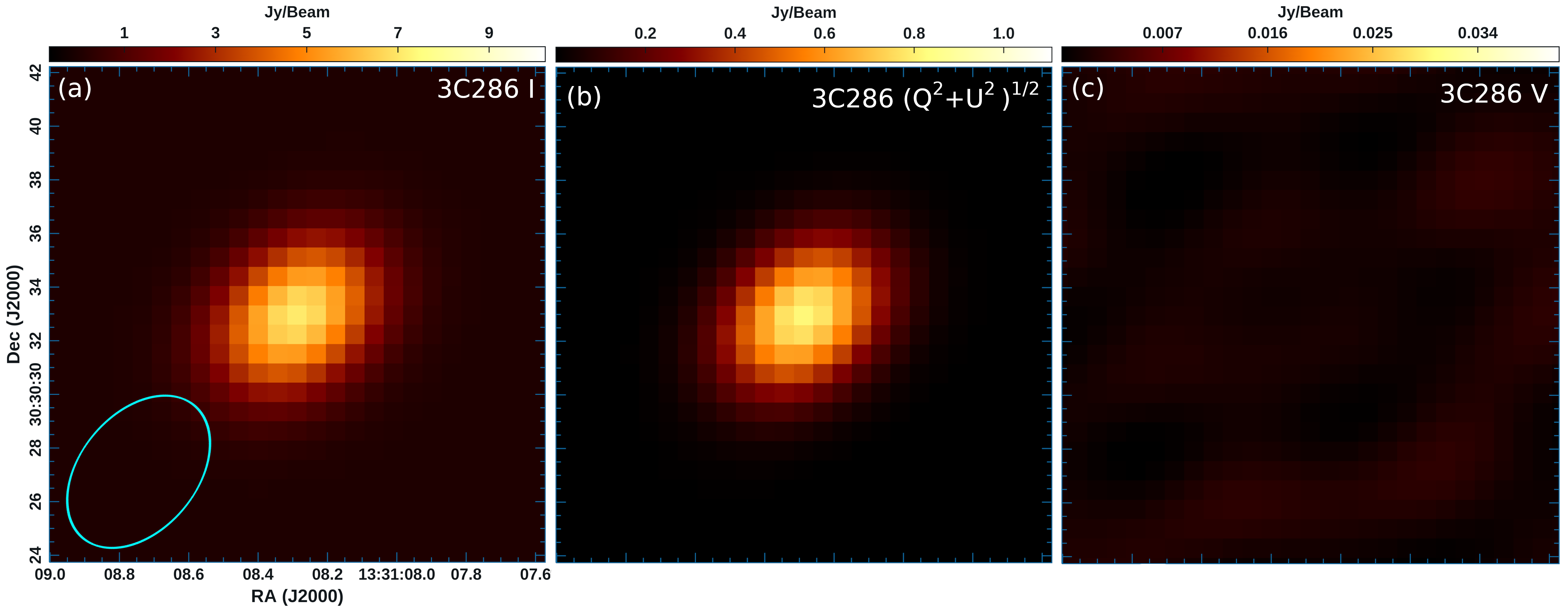}
    \caption{(a): Stokes $I$ emission of the flux calibrator 3C286. (b): Stokes $\sqrt{Q^2 + U^2}$ emission of the flux calibrator 3C286. (c): Stokes $V$ image of the flux calibrator 3C286. The cyan ellipse represents the beam.}
    \label{fig:flux}
\end{figure}

\begin{figure}[H]
    \centering
    \includegraphics[width=\textwidth]{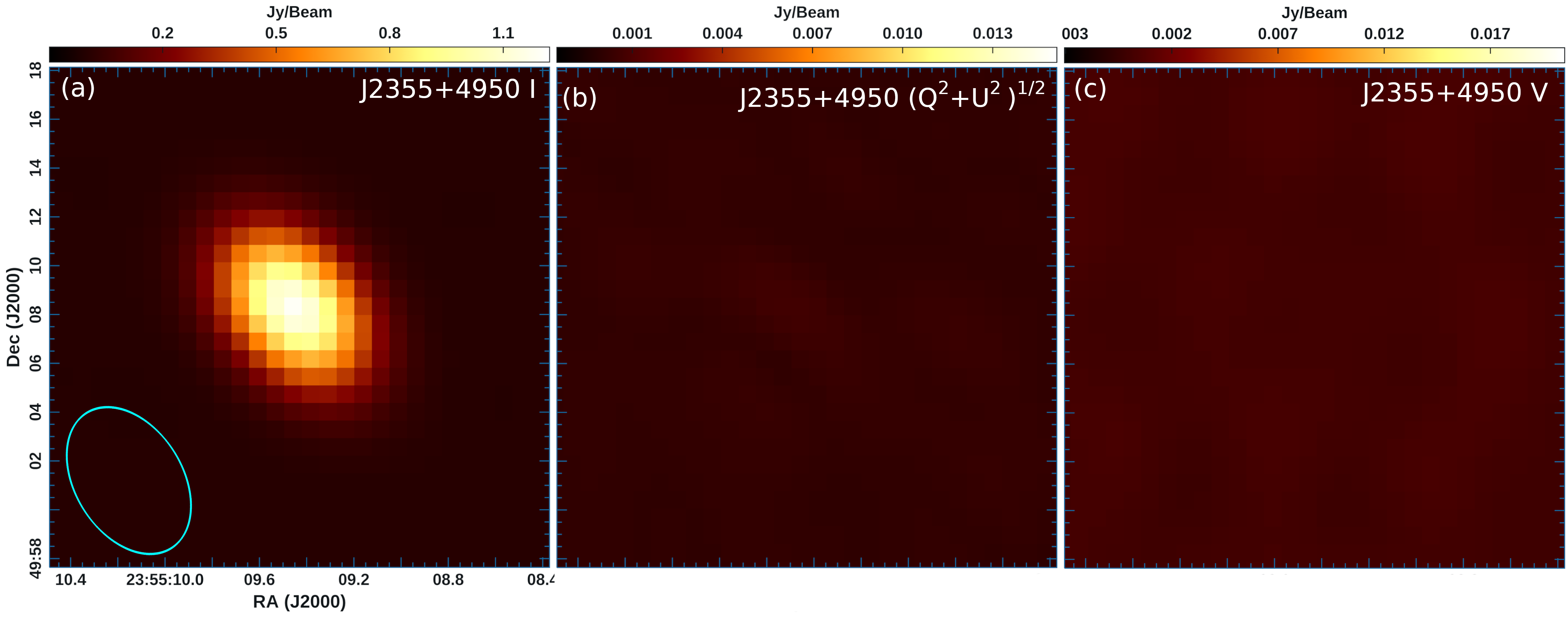}
    \caption{(a): Stokes $I$ emission of the polarization calibrator J2355+4950. (b): Stokes $\sqrt{Q^2 + U^2}$ image of J2355+4950. (c): Stokes $V$ image of J2355+4950. The cyan ellipse represents the beam.}
    \label{fig:pol}
\end{figure}

In order to ensure that the Stokes $V$ emission is not due to leakage from the Stokes $I$ image or due to instrumental polarization, we have carried out multiple checks. Firstly, if this emission is due to leakage, then all the other radio sources with similar flux density as I18162 (e.g. S9, S16) or higher flux density (e.g. S2) should also have associated $V$ emission in the image. However, we do not detect CP towards S9, S16, etc. We do detect CP towards the bright source S2, located to the west of S9, which is outside the region displayed in the maps, whose nature is unknown. Secondly, we checked the polarization calibration to ensure the instrumental polarization is negligible and below the rms level. The primary calibrator 3C286 and the polarization calibrator J2355+4950 have been imaged to scrutinize for leakage emission. The images of these two calibrators can be found in Figs.~\ref{fig:flux} \& \ref{fig:pol}. 3C286 is linearly polarized and has a polarization fraction in the range $13-16\%$ in the VLA C band. J2355+4950, on the other hand, is unpolarized. Towards 3C286, the LP fraction is $\sim14\%$ consistent with expectations at these frequencies, while no emission is detected in the Stokes $V$ image. Towards J2355+4950, no emission is detected in Stokes $Q$, $U$ and $V$ bands.

Finally, we check for the beam squint effect, which could result from the offset between right circularly polarized (RCP) and left circularly polarized (LCP) receivers in the VLA antennas \citep{2008A&A...486..647U} leading to spurious emission patterns in Stokes parameters $Q$, $U$, $V$ images, as well as fluctuation patterns in flux density over time \citep{2017AJ....154...56J}. The most convenient way to see the effects of beam squint and to test the efficacy of the correction is to examine a plot of the time-variable Stokes $V$ amplitudes averaged over spectral channels \citep{2008A&A...486..647U}. The instrumental polarization is fixed to the antenna and will rotate with a parallactic angle, causing a time-variable Stokes $V$ signal in the averaged data, which varies between positive and negative values. We averaged three scans and imaged each one to check for flux variations with time in our data for both days (21 and 22 December 2018). The flux remained constant throughout the scans for both days, neglecting the possibility of the beam squint effect. In addition to this, the effect can introduce artifacts due to the variable Stokes $V$ signal induced by the Earth’s rotation of the (squinted) primary beam response with respect to the sky. No artifacts were observed in our maps, which further validates the calibration. All these checks suggest that the calibration is of high quality, thereby lending credibility to the detected emission.

\onecolumngrid


\end{document}